\documentclass[conference]{IEEEtran}
\IEEEoverridecommandlockouts
\usepackage{cite}
\usepackage{amsmath,amssymb,amsfonts}
\usepackage{algorithmic}
\usepackage{graphicx}
\usepackage{textcomp}
\usepackage{xcolor}
\usepackage{hyperref}
\hypersetup{
    colorlinks=true,
    linkcolor=black,
    urlcolor=blue,
    citecolor=black
    }
\def\BibTeX{{\rm B\kern-.05em{\sc i\kern-.025em b}\kern-.08em
    T\kern-.1667em\lower.7ex\hbox{E}\kern-.125emX}}
\begin{document}

\title{Machine Learning Based Forward Solver: An Automatic Framework in gprMax 
}

\author{\IEEEauthorblockN{Utsav Akhaury\IEEEauthorrefmark{1}, Iraklis Giannakis\IEEEauthorrefmark{2}, Craig Warren\IEEEauthorrefmark{3}, and Antonios Giannopoulos\IEEEauthorrefmark{4}\\ \\}
\IEEEauthorblockA{\IEEEauthorrefmark{1}Laboratoire d’astrophysique, École Polytechnique Fédérale de Lausanne (EPFL), Switzerland}
\IEEEauthorblockA{\IEEEauthorrefmark{2}School of Geosciences, University of Aberdeen, Aberdeen, United Kingdom}
\IEEEauthorblockA{\IEEEauthorrefmark{3}Department of Mechanical and Construction Engineering,
Northumbria University,
Newcastle, United Kingdom}
\IEEEauthorblockA{\IEEEauthorrefmark{4}School of Engineering, The University of Edinburgh, Edinburgh, United Kingdom}
f2016684p@alumni.bits-pilani.ac.in, iraklis.giannakis@abdn.ac.uk, craig.warren@northumbria.ac.uk, a.giannopoulos@ed.ac.uk
% <-this % stops an unwanted space
}

\iffalse
\author{\IEEEauthorblockN{Utsav Akhaury}
\IEEEauthorblockA{\textit{Laboratoire d’astrophysique}\\
%\textit{Faculty of Fundamental Problems of Technology} \\
\textit{École Polytechnique Fédérale de Lausanne (EPFL)}\\
Lausanne, Switzerland \\
f2016684p@alumni.bits-pilani.ac.in}
\and
\IEEEauthorblockN{Iraklis Giannakis}
\IEEEauthorblockA{\textit{School of Geosciences} \\
\textit{University of Aberdeen}\\
Aberdeen, United Kingdom \\
iraklis.giannakis@abdn.ac.uk}
\and
\IEEEauthorblockN{Craig Warren}
\IEEEauthorblockA{\textit{Department of Mechanical and Construction Engineering} \\
\textit{Northumbria University}\\
Newcastle, United Kingdom \\
craig.warren@northumbria.ac.uk}
\and
\IEEEauthorblockN{Antonios Giannopoulos}
\IEEEauthorblockA{\textit{School of Engineering} \\
\textit{The University of Edinburgh}\\
Edinburgh, United Kingdom \\
a.giannopoulos@ed.ac.uk}
}
\fi
\maketitle

\begin{abstract}
  General full-wave electromagnetic solvers, such as those utilizing the finite-difference time-domain (FDTD) method, are computationally demanding for simulating practical GPR problems. We explore the performance of a near-real-time, forward modeling approach for GPR that is based on a machine learning (ML) architecture. To ease the process, we have developed a framework that is capable of generating these ML-based forward solvers automatically. The framework uses an innovative training method that combines a predictive dimensionality reduction technique and a large data set of modeled GPR responses from our FDTD simulation software, gprMax. The forward solver is parameterized for a specific GPR application, but the framework can be extended in a straightforward manner to different electromagnetic problems. \\
\end{abstract}


\section{Introduction}
Recent progress in the field of GPR simulations \cite{b1} enabled research on developing advanced modelling tools based on big data and machine learning (ML). In this paper, we explore the concept of a near-real-time, forward modeling approach for GPR that is based on ML architectures. The initial idea for our ML approach was developed in \cite{b2} and further enhanced in \cite{b3}. Our approach greatly eases the ML-based framework generation by introducing the Random Parameter Generation feature in gprMax \cite{b4}, our bespoke FDTD simulation software. Using this feature, we generate a large number of GPR models with randomly varying parameters. Subsequently, we reduce the dimensionality of these data similar to \cite{b3}, and finally, we compare the performance of different ML regressors that take as inputs these random parameters and the compressed A-Scans of the models for training. The ultimate goal is to predict the output responses of GPR models depending on the selected parameters used as inputs by the user. The whole process is automatic, and although the training is time-consuming and computationally demanding, the final output is a near real-time forward solver capable of predicting the resulting A-Scan subject to the given inputs. 

A fast numerical solver can greatly accelerate full-waveform inversion (FWI), since the most computationally intensive module of the latter is the emended numerical modelling in each iteration. This would make FWI commercially appealing and attainable using minimum computational resources without the need for high performance computing. \\

\section{Dataset Generation}

\subsection{Random Parameter Generation -  New Module in gprMax}
This new feature facilitates the generation of random parameters for a specific gprMax model and uses the \emph{numpy.random} module from NumPy \cite{b5} as backend. It allows the user to specify the Probability Distribution Function (PDF) from which these random parameters are drawn. Subsequently, for every numerical parameter defining the GPR scenario, the user must enter two values (in pairs) that define the specified PDF's parameters. The following convention is followed to activate the random parameter generation mode:
\begin{itemize}
\item \emph{\#command\_name: distr parameter\_1.1 parameter\_1.2 parameter\_2.1 parameter\_2.2 parameter\_3.1 parameter\_3.2 ...}
\end{itemize}

\emph{distr} specifies the PDF from which random numbers are drawn. For example, to specify a uniform distribution, the user must enter \emph{u}, and the two subsequent values (\emph{parameter\_x.1 \& parameter\_x.2})  entered would define the lower \& upper bounds for the $x^{th}$ parameter. Note that if \emph{parameter\_x.1 = parameter\_x.2}, then the random number generation is skipped and the constant value is used.

As a quick demonstration, the following command creates a material called \emph{my\_sand} which has a relative permittivity $\epsilon_r$ drawn from a uniform distribution within the range $[2,5]$, a conductivity of $\sigma = 0.01 S/m$, $\mu_r = 1$ and $\sigma_* = 0$.
\begin{itemize}
\item \emph{\#material: u 2 5 0.01 0.01 1 1 0 0 my\_sand}
\end{itemize}

In case the generated random parameter exceeds the model domain bounds, it is automatically constrained to fit inside the domain, which ensures that the FDTD execution is not stopped midway. Finally, all the randomly generated parameters for every GPR model generated are saved to a pickle file. We automatically compress the pickle file by removing redundant features (i.e. those features that have a constant value for every GPR model generated) in order to improve the training efficiency. 

\subsection{Metal Cylinder buried in a Dielectric Medium}

\begin{figure}[htbp]
\centerline{\includegraphics[scale=0.075]{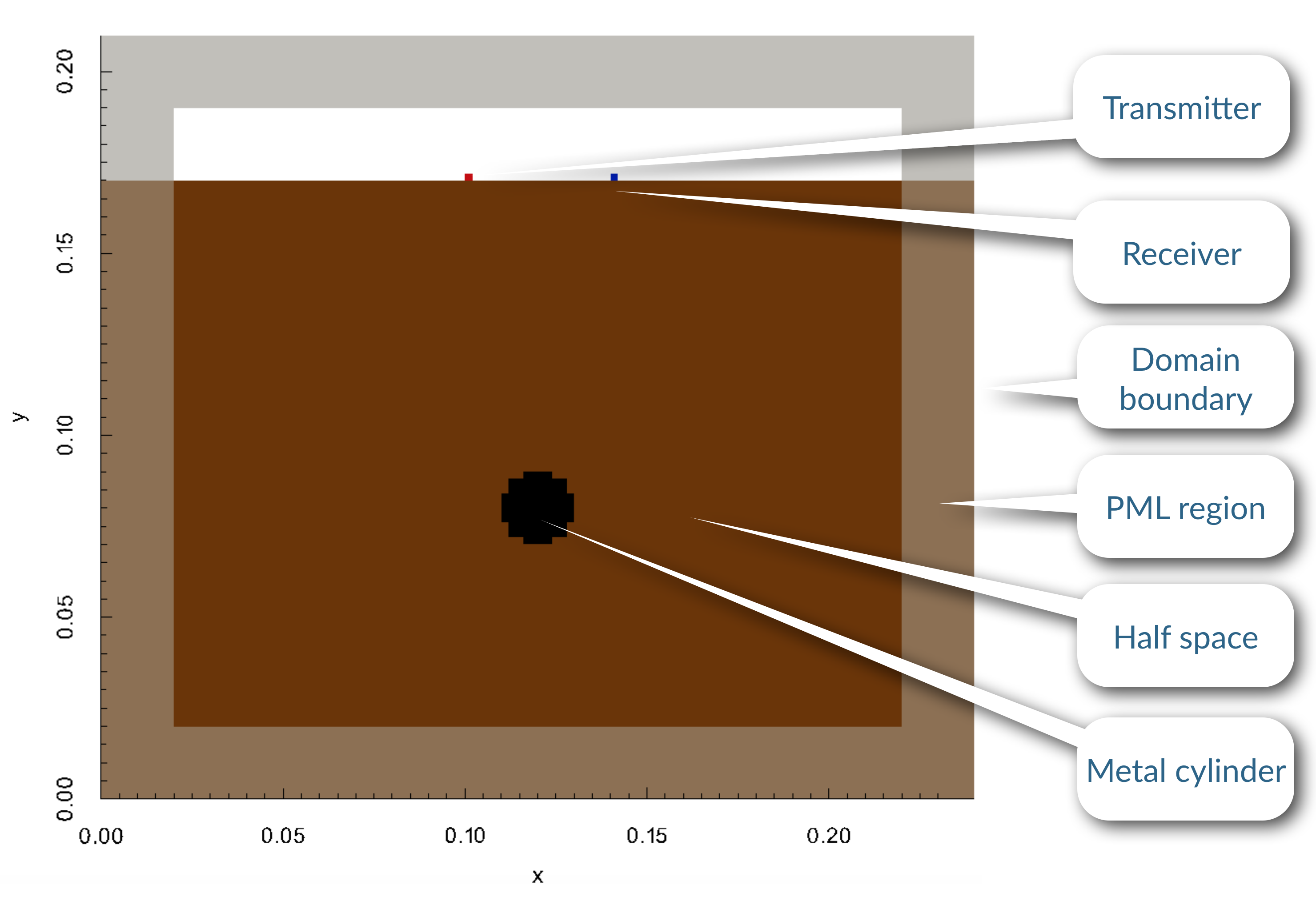}}
\caption{Geometry of a sample GPR model generated.}
\label{model_geo}
\end{figure}

Using the Random Parameter Generation feature detailed above, we generated a dataset of 6250 2D GPR models for a metal cylinder buried in a dielectric half-space. We then randomly split the dataset into train-test subsets - 5000 models for training and 1250 for testing. The geometry of the simple 2D scenario is straight-forward and an image from the geometry view is shown in Fig. \ref{model_geo}. The transparent region around the boundary of the domain depicts the PML (Perfectly matched layer) region. The line source used for the model's excitation is polarized along the $z$ direction, and is excited using a Ricker waveform with amplitude of one and centre frequency of 1.5 GHz. Table \ref{tab1} summarizes the parameters that remain constant for every model.

In each of these models, we vary the following defining parameters, which are drawn from a Uniform Distribution. 

\begin{itemize}
\item The cylinder radius
\item The depth at which the cylinder is placed in the dielectric medium
\item The electromagnetic permittivity ($\epsilon_r$) of the dielectric half-space
\end{itemize}
These are summarized in Table \ref{tab2}.

The electromagnetic response (output A-Scans) of all these models are simulated using the gprMax software and saved along with all the randomly generated parameters. These data-pairs would eventually be fed to the ML scheme. Note that since the line source is aligned along the $z$ direction, only the $E_z$ component of the electric field is non-zero. And hence, we only use the $E_z$ component for training. 

\begin{table}[htbp]
\caption{Constant Model Parameters}
\begin{center}
\begin{tabular}{|c|c|}
\hline
\textbf{\textbf{Parameter}} & \textbf{\textbf{Values}} \\
\hline
Domain Bounds $(x,y,z)$ & 0-240 mm, 0-210 mm, 0-2 mm \\
Dielectric half-space bounds $(x,y,z)$ & 0-240 mm, 0-170 mm, 0-2 mm \\
Spatial Discretization $(dx,dy,dz)$ & 2 mm, 2 mm, 2 mm \\
Time Window & 3 ns \\
Source Location $(x,y,z)$ & 100 mm, 170 mm, 0 mm \\
Receiver Location $(x,y,z)$ & 140 mm, 170 mm, 0 mm \\
$\sigma$ - dielectric half-space & 0 \\
$\sigma_*$ - dielectric half-space & 0 \\
$\mu_r$ - dielectric half-space & 1 \\
\hline
\end{tabular}
\label{tab1}
\end{center}
\end{table}

\begin{table}[htbp]
\caption{Randomly Varying Parameters}
\begin{center}
\begin{tabular}{|c|c|c|}
\hline
\textbf{\textbf{Parameter}} & \textbf{\textbf{Lower Bound}}& \textbf{\textbf{Upper Bound}} \\
\hline
Cylinder Radius (r) & 5 mm & 20 mm \\
Cylinder Depth & 20 mm & 140 mm \\
$\epsilon_r$ - dielectric half-space & 4 & 8 \\
\hline
\end{tabular}
\label{tab2}
\end{center}
\end{table}

\section{Dimensionality Reduction}
To increase the efficiency of our ML-based forward solver, we pre-process our data in a suitable manner. To that end, we reduce the dimensionality of the A-Scans before feeding them to the ML model. It is observed that the A-Scans can effectively be represented using much lesser number of components without any significant loss in signal quality. We compared the performance of two popular algorithms, namely Principle Component Analysis (PCA) and Singular Value Decomposition (SVD), which give us very similar results in terms of degree of compression. 

In either case, we start by using 10 components and computing the Normalized Mean Squared Error (NMSE) between the original and reconstructed signals. We then iteratively increase the number of components used for representing the signal until the difference between successive NMSEs falls below a certain threshold ($10^{-12}$). We noticed that most of the signal information is captured by the first 10 to 15 components. Obviously, adding more components would further minimize the NMSE. However, even with around 30 components, we get an NMSE $\approx 10^{-12}$ on 1250 test samples, which is sufficiently low for our purpose.

\subsection{Principle Component Analysis (PCA)}
Principal component analysis (PCA) is among the oldest methods for reducing the dimensionality of datasets \cite{b6}, and it does so by increasing interpretability while minimizing information loss. It performs this operation by creating new uncorrelated variables that consecutively maximize variance. These new variables are called the principal components. Finding these principle components reduces to solving an eigenvalue/eigenvector problem, and the new variables are defined by the dataset at hand, not a priori, thus rendering PCA an adaptive data analysis technique. PCA has already been successfully applied to GPR data for clutter reduction in \cite{b7} - \cite{b8}, and more precisely for dimensionality reduction in \cite{b9}. 

\begin{figure}[htbp]
\centerline{\includegraphics[scale=0.25]{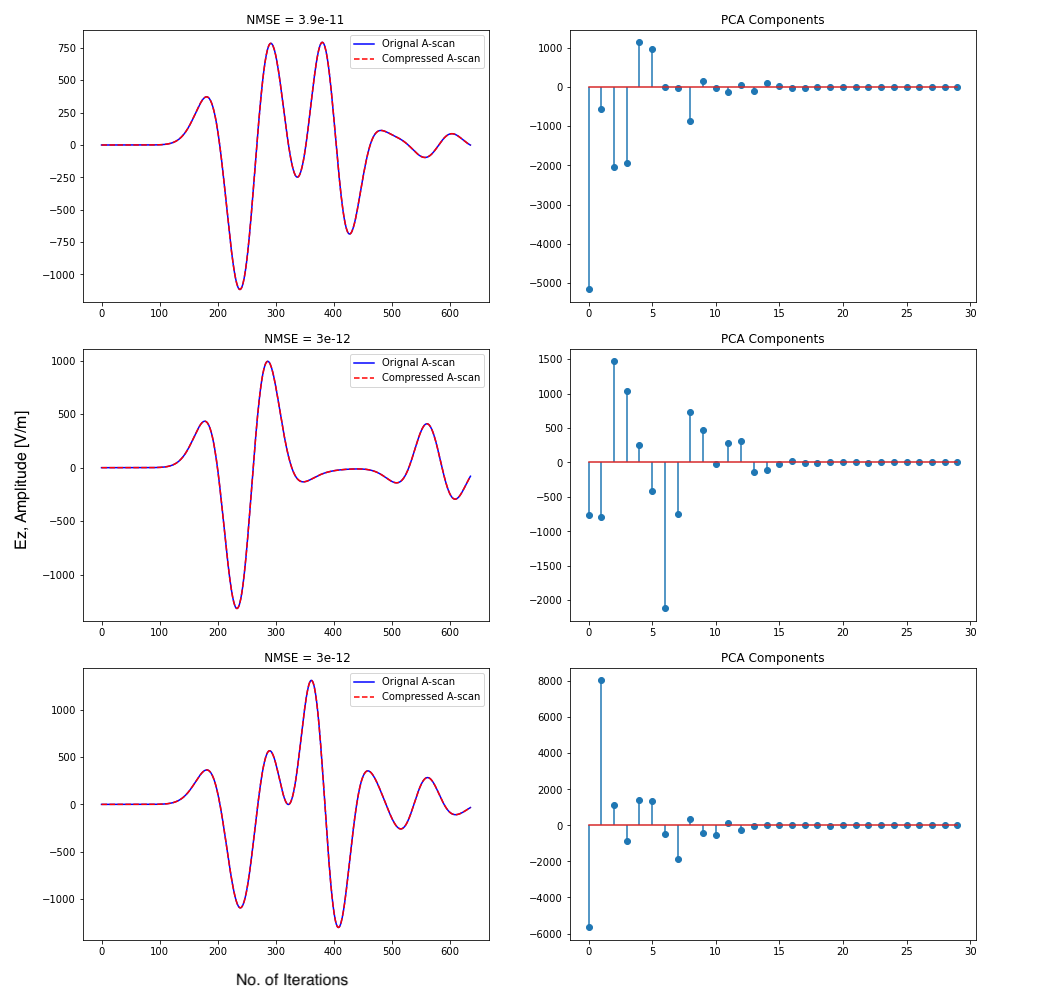}}
\caption{Dimensionality Reduction using PCA. The dimension is reduced from 637 to 30, and yet the reconstruction is in excellent agreement with the original signal. Notice that most of the signal information is captured by the first 10 components, which are much larger in magnitude compared to the ones that follow.}
\label{pca}
\end{figure}

Fig. \ref{pca} shows a few comparison plots between the A-Scans reconstructed from their compressed representations (by applying inverse PCA transform) and the original A-Scans. 

\subsection{Singular Value Decomposition (SVD)}
Singular value decomposition (SVD) is a factorization method of a real or complex matrix that extends the concept of the eigen-decomposition of a square normal matrix with an orthonormal eigenbasis to any $m \times n$ matrix \cite{b10}. Truncated SVD is a quicker and more compact type of SVD in which the data matrix is truncated and the rest of it is discarded. This can be much quicker and more economical than simple SVD.

\begin{figure}[htbp]
\centerline{\includegraphics[scale=0.25]{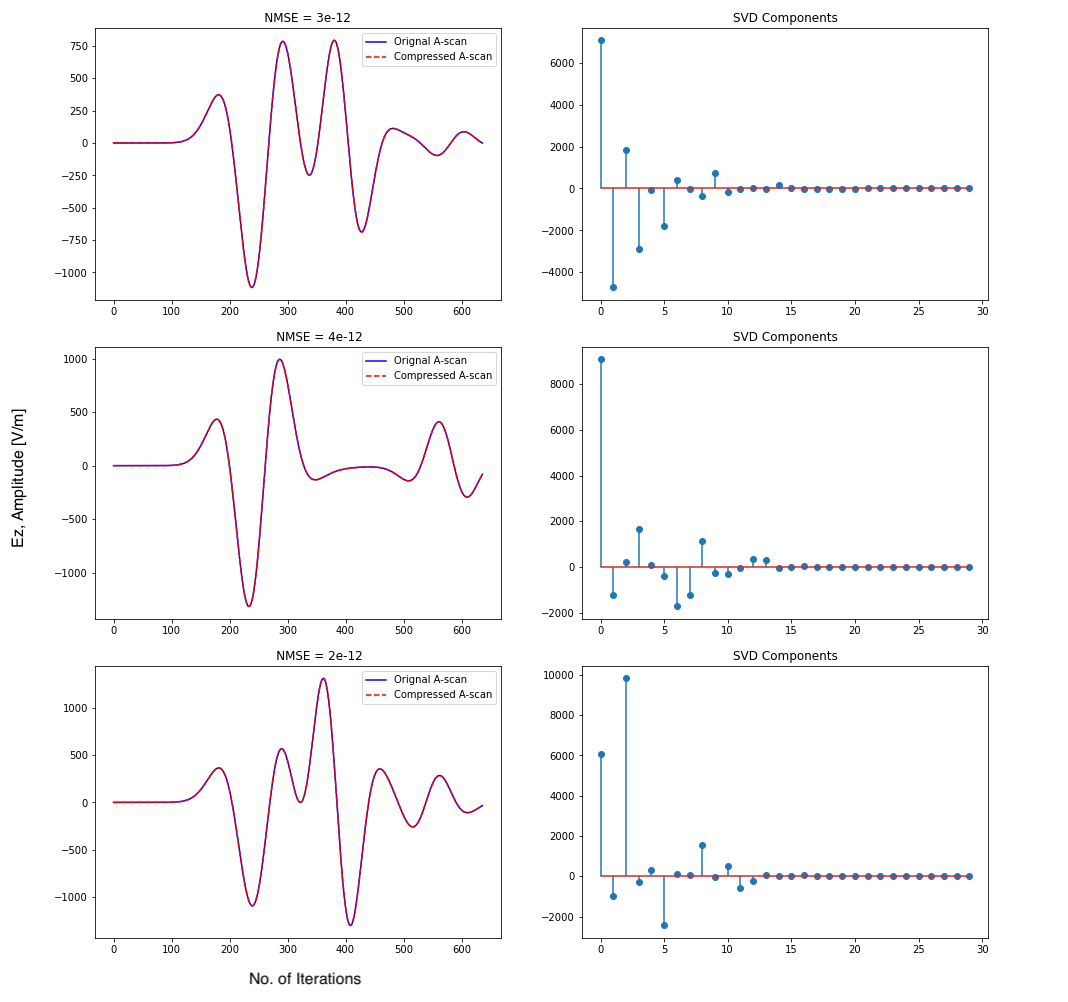}}
\caption{Dimensionality Reduction using TruncatedSVD. The dimension of the A-Scans is reduced from 637 to 30.}
\label{svd}
\end{figure}

For our work, we use scikit-learn's TruncatedSVD transformer. As opposed to PCA, this estimator does not center the data before performing SVD, which implies that it can efficiently handle sparse matrices. Fig. \ref{svd} shows a few comparison plots between the A-Scans reconstructed from their compressed representations (by applying inverse SVD transform) and the original A-Scans.

\section{Machine Learning based Forward Solver}

Once we have the compressed representations of the output responses of our GPR models, we feed them to our ML-based model for training. Supervised machine learning will unravel the physical causal relationship between the inputs and the resulting trace, and effectively approximate the underlying physical laws of the given problem. To that end, we compare the performance of a few popular ML regressors, such as Random Forest \cite{b11} and XGBoost \cite{b12}, that we extend to perform multi-variate multi-output regression. We had also evaluated the performance of other regressors such as Support Vector Machines (SVM) \cite{b13}, etc. However, they do not even come close to the accuracy achieved by Random Forest or XGBoost, and hence, we only focus on these two algorithms in our work. 

The ML solver outputs the compressed form of the output A-Scan, which can be reconstructed by inverting the same dimensionality reduction algorithm that was used to compress the training data.

\subsection{Random Forest}
Random Forest \cite{b11} is a supervised machine learning algorithm that is constructed using decision trees \cite{b14} and invokes the concept of ensemble learning \cite{b15}, a technique that combines predictions from multiple ML algorithms to make a more accurate prediction compared to a single model. As an improvement to the decision tree algorithm, it reduces overfitting to datasets and increases precision. It is a bagging-based algorithm in which only a subset of features is selected at random. The maximum tree depth, which is a tunable input parameter, controls the tendancy of the model to overfit to the input data. In case of regression, Random Forest returns the mean or average prediction of the individual trees. It can handle large datasets with ease but does not perform well on very sparse data \cite{b16}. The idea of using Random Forest regression for GPR-based predictions has been introduced in \cite{b17}. 

\begin{figure}[htbp]
\centerline{\includegraphics[scale=0.39]{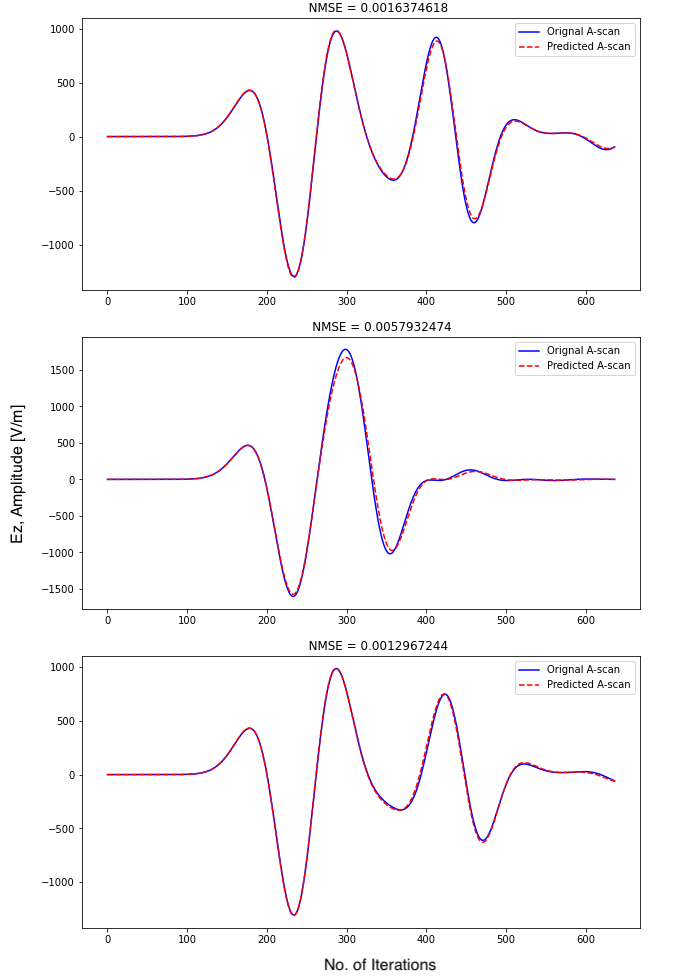}}
\caption{Output predictions for Random Forest.}
\label{rf}
\end{figure}

We evaluate the performance of our trained Random Forest model on our test dataset. A few output predictions using Random Forest are shown in Fig. \ref{rf}.

\subsection{XGBoost}
\begin{figure}[htbp]
\centerline{\includegraphics[scale=0.39]{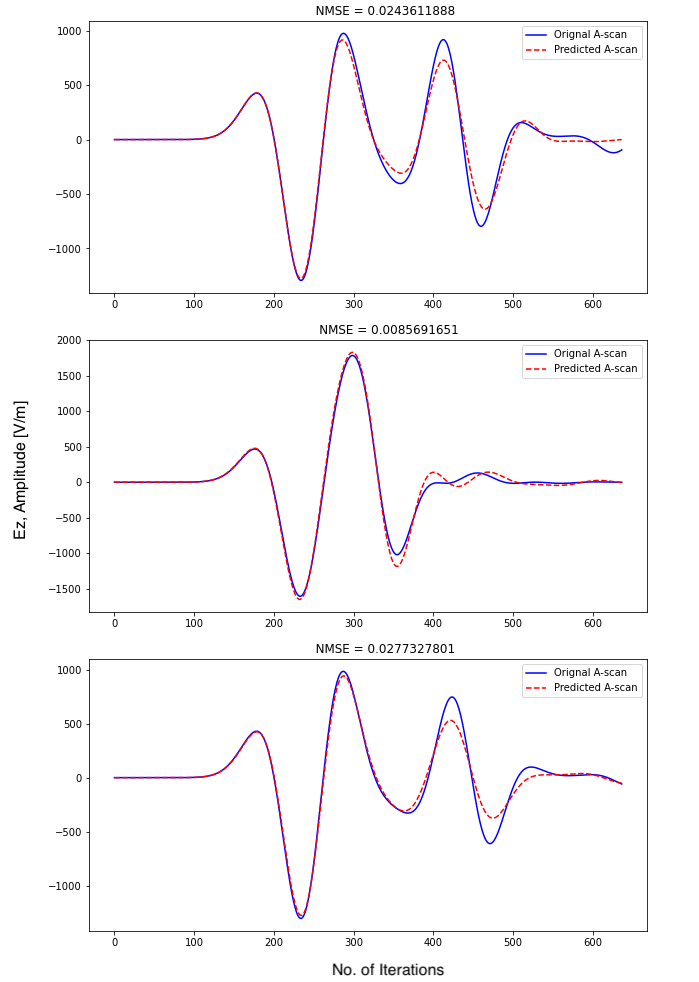}}
\caption{Output predictions for XGBoost.}
\label{xgb}
\end{figure}

XGBoost \cite{b12}, or Extreme Gradient Boosting, is also a decision-tree-based ensemble Machine Learning algorithm that uses a gradient boosting framework. It builds upon the concept of gradient boosting algorithm and provides a more efficient implementation by invoking parallelization, tree pruning, and hardware optimization. For most regression problems, XGBoost provably has the best combination of prediction performance and processing time compared to other algorithms. A few output predictions using XGBoost are shown in Fig. \ref{xgb}.

\subsection{Chain Regression}
A Chain Regressor is a multi-label model that arranges regressions into a chain. Every model makes a prediction within the order determined by the chain using all of the available features provided to the model and the predictions of models that precede in the chain. Chain Regressors are suitable where multi-output regression is required and can improve the training performance, as shown in \cite{b18}. However, they also take significantly longer to train. 

\begin{figure}[htbp]
\centerline{\includegraphics[scale=0.39]{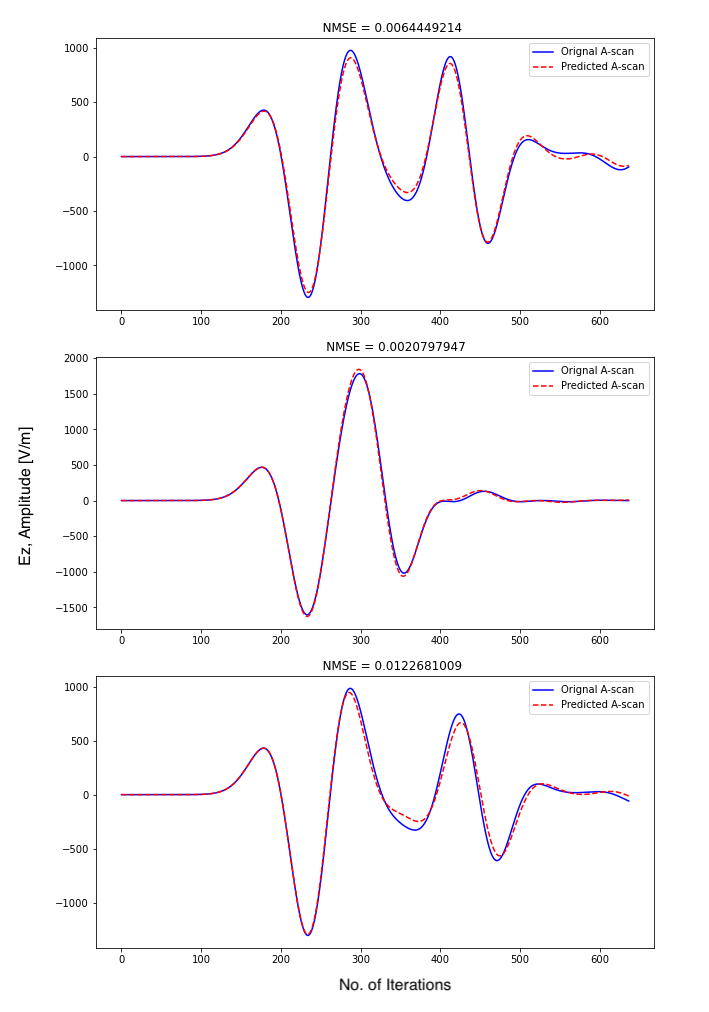}}
\caption{Output predictions for XGBoost with Chain Regression.}
\label{xgbchain}
\end{figure}

We test the performance of Random Forest and XGBoost arranged in regression chains. However, it is observed that there is no visible improvement when the two cases are applied to our test dataset. On the contrary, it leads to a degradation in performance. A few output predictions using Chain Regression for XGBoost are shown in Fig. \ref{xgbchain}. 

\section{Results and Comparison}

Table \ref{tab3} shows a comparison of the performance of the various ML schemes we have tested. We applied all the methods discussed above to our test dataset of 1250 GPR models. All algorithms were tested on a 2.3 GHz Quad-Core Intel Core i7 processor. For a quantitative comparison, we computed the Normalized Mean Squared Error (NMSE) between the predicted and the simulated ground-truth output responses (that were obtained using gprMax). Note that a lower value of NMSE implies better performance. Alongside, we also noted the time taken by the algorithm for training.

Random Forest achieves the least NMSE on our test dataset, followed by XGBoost. As mentioned earlier, introducing Chain Regression does not show any improvement compared to the original multi-output regressors. Furthermore, it significantly consumes more time during training.

\begin{table}[htbp]
\caption{Performance Comparison}
\begin{center}
\begin{tabular}{|c|c|c|}
\hline
\textbf{\textbf{Method}} & \textbf{\textbf{NMSE}}& \textbf{\textbf{Training Time}} \\
\hline
Random Forest & 0.0182 & 2.8s\\
Random Forest + Chain Regression & 0.0809 & 179.5s \\
XGBoost & 0.0285 & 8.8s\\
XGBoost + Chain Regression & 0.1011 & 17.4s\\
\hline
\end{tabular}
\label{tab3}
\end{center}
\end{table}

\section{Conclusion}
We successfully implemented a framework for generating ML-based forward solvers in gprMax, our FDTD simulation software. This was greatly eased by introducing the Random Parameter Generation feature. We observed that for our test dataset, Random Forest gives us the best performance, taking into consideration both, the NMSE between the ground truth \& output predictions, and the training time. Ultimately, gprMax gives the user the flexibility to train \& test the performance of any other suitable ML model to better fit a specific dataset. Furthermore, as shown in \cite{b3}, the performance of neural networks \cite{b19} could also be investigated on our dataset in detail, which remains the future scope of this project. The ML solver can also easily handle 3D geometries, in which case, there would only be a greater number of parameters. The underlying principle remains the same.

\section*{Acknowledgment}
The project was funded via the Google Summer of Code (GSoC) 2021 programme. GSoC initiative is a global program focused on bringing student developers into open source software development. The source code for this project can be found at \url{https://github.com/gprMax/gprMax/pull/294}.

\end{document}